\documentclass{article}

\usepackage{arxiv}
\usepackage[utf8]{inputenc} 
\usepackage[T1]{fontenc}    
\usepackage{hyperref}       
\usepackage{url}            
\usepackage{booktabs}       
\usepackage{amsfonts}       
\usepackage{nicefrac}       
\usepackage{microtype}      
\usepackage{lipsum}
\usepackage{natbib}
\setcitestyle{numbers}
\usepackage{amsmath,amssymb}
\usepackage{changepage}
\usepackage{graphicx}

\usepackage{color}
\usepackage{subfigure}
\usepackage{bm}
\usepackage{soul}

\usepackage{xcolor}

\newcommand{\neww}[1]{{{#1}}}

\newcommand{\new}[1]{{{#1}}}

\title{Intermittent percolation and the scale-free distribution of
  vegetation clusters}

\author{
  Paula Villa Mart{\'\i}n \\
  Okinawa Institute of Science and Technology Graduate University, \\
  Biological Complexity Unit, Onna, Okinawa 904-0495, Japan \\
   \AND
Virginia Dom{\'\i}nguez-Garc{\'\i}a \\
  ISEM, CNRS, Univ Montpellier, EPHE, IRD, 34095 Montpellier, France \\
  \AND
  Miguel A. Mu\~noz \\
  Departamento de Electromagnetismo y F{\'i}sica de la Materia \\
  and Instituto Carlos I de F{\'i}sica Te{\'o}rica y
  Computacional, \\
  Universidad de Granada, Granada, Spain \\
  \texttt{mamunoz@onsager.ugr.es} \\
}

\begin{document}
\maketitle

\begin{abstract}
  Understanding the causes and effects of spatial vegetation patterns
  is a fundamental problem in ecology, especially because these can be
  used as early predictors of catastrophic shifts such as
  desertification processes.  Empirical studies of the vegetation
  cover in some areas such as drylands and semiarid regions have
  revealed the existence of vegetation patches of broadly diverse
  sizes. In particular, the probability distribution of patch sizes
  can be fitted by a power law, i.e. vegetation patches are
  approximately scale free up to some maximum size.  Different
  explanatory mechanisms, such as plant-plant interactions and
  plant-water feedback loops have been proposed to rationalize the
  emergence of such scale-free patterns, yet a full understanding has
  not been reached.  Using a simple model for vegetation dynamics, we
  show that environmental temporal variability --a well-recognized
  feature of semiarid environments-- promotes in a robust way
  (i.e. for a wide range of parameter values) the emergence of
  vegetation patches with broadly distributed cluster
  sizes. \neww{Furthermore, this result is related to a percolation
    phenomenon that occurs in an intermittent or fluctuating way. The
    model also reveals that the power-law exponents fitting the tails
    of the
    probability distributions depend on the overall vegetation-cover
    density,} in agreement with empirical observations. This supports
  the idea 
  that environmental variability plays a key role in the formation of
  scale-free vegetation patterns. From a practical viewpoint, this may
  be of importance to predict the effects that changes in
  environmental conditions may have in real ecosystems. From a
    theoretical side, our study sheds new light on a novel type of
    percolation phenomena occurring under temporally-varying external
    conditions, \neww{that still needs further work to be fully characterized.}
\end{abstract}

\section{Introduction}

Most species are typically distributed in space neither uniformly nor
randomly, but forming patterns of spatial aggregation
\cite{Condit,Staver}.  Understanding the causes and effects of such
spatial aggregation is one of the most fundamental problems in
theoretical ecology
\cite{Levin,Durrett-Levin,Legendre,Manrubia,Sole,Sole+Bascompte,Azaele,JSP}.  An
archetypical example of vegetation patterns is found in arid and
semiarid environments, which are characterized by the presence of
vegetation patches, organized in either regular or irregular patterns,
even in seemingly spatially homogeneous environments.
\cite{Tarnita-JA,Regular,semiarid1,semiarid2,semiarid3,Lejeune}.  A
large number of mechanisms have been proposed to explain the origin of
such vegetation patterns. These include, among others, the interplay
between competition and facilitation processes in plant-plant
interactions as well as plant-water feedbacks
\cite{Scanlon2007,Kefi2007,Manor-Shnerb-2008a,Manor-Shnerb-2008b,Manor-Shnerb-2009},
but a full understanding of how spatial patterns emerge is still
lacking. Importantly, even if this idea remains controversial,
variations in the shape of vegetation patterns as well as in the
density cover have been proposed as early-warning indicators of
catastrophic shifts, such as desertification
\cite{Scheffer2001,Rietkerk2004,Kefi2007,Kefi2011,Kefi-Nature2017}.
Thus, in summary, understanding, characterizing, and eventually
controlling vegetation patterns is a problem of outmost theoretical
and practical relevance.

While some of the patterns formed by vegetation, such as stripes, have
a characteristic size scale, ground-breaking empirical studies showed
a decade ago that vegetation aggregates in semiarid environments lack
a well-defined characteristic scale (i.e. vegetation patches of
greatly diverse sizes can be observed in nearby areas)
\cite{Scanlon2007,Kefi2007}.  More specifically, in this latter
situation patch-size distributions can be fitted by power-laws \cite{Manrubia,Azaele}, the
hallmark of scale-free systems
\cite{Newman-powerlaws,Mitzenmacher,Sornette-book,RMP}.  From a
theoretical perspective, scale-free cluster-size distributions are
often the fingerprint of critical points, at the edge of a phase
transition
\cite{Binney,Henkel-book,Stauffer-Aharony,Christensen-book,GG}. An
archetypical example of a critical point, paradigmatic to rationalize
scale-free clusters, is ``standard'' or ``isotropic'' percolation.

In the simplest percolation model, the sites of a given lattice (which
represents the physical space) can be in two possible states: either
empty or occupied.  In the percolating phase, i.e. for large
occupation probabilities, the largest aggregate of occupied sites
spans the whole system; on the other hand, in the non-percolating
phase, i.e. for low occupation probabilities, only finite-size
clusters exist.  In between these two phases --right at the
percolation threshold or critical point-- clusters of all possible
sizes --from tiny to extremely large ones-- emerge, generating a
power-law (scale-free) distribution of cluster sizes; this has
well-known universal critical exponents and characteristic scaling
features \cite{Stauffer-Aharony,Christensen-book}.

If the ecological processes giving rise to scale-free vegetation
patches in semiarid ecosystems would generate --for some unknown
reason-- clusters such as those close to a percolation phase
transition \cite{Roy2003,Pascual2005}, their observed scaling
exponents should be universal. \neww{In other words, robust power-law
  exponents, not depending e.g. on vegetation density, should be
  observed} \cite{Binney,Henkel-book}.  However, contrary to this
expectation, the empirically found distributions of patch sizes can be
fitted by power laws whose exponents are system dependent and change
with the level of vegetation coverage in the analyzed area
\cite{Scanlon2007,Kefi2007,Kefi-Nature2017} (see, however,
\cite{Staver}). This variability in scaling exponents suggests a lack
of universality and sheds doubts on the --exceedingly naive-- alleged
connection with standard/isotropic percolation and universality,
calling for more elaborated theoretical approaches to understand the
dynamics and patterns of these ecological communities.

Diverse relatively-simple dynamical models have been proposed to
account for the empirical finding of scale-free vegetation patches
\cite{Scanlon2007,Kefi2007,Manor-Shnerb-2008a,Manor-Shnerb-2008b,Manor-Shnerb-2009}.
Some of these include three possible states to describe the system
at each  spatial location:
occupied, empty, and disturbed/degraded (i.e. unavailable to be
colonized by vegetation)  \cite{Kefi2007,Roy2003}.
In such three-state models, an external control parameter (which can
be interpreted as the grazing pressure, the yearly rainfall, fire
etc.)  regulates the level of vegetation cover under stationary
conditions. As it is varied --e.g. to model transiently harsher or
milder environmental conditions-- a phase transition from the
vegetated to the deserted state may appear
\cite{Roy2003,Kefi2007,Kefi2011}.  Remarkably, it has been found
--relying mostly on computational analyses-- that cluster-size
distributions resembling power-laws can be observed in some of these
models for a wide range of external conditions (a broad set of
parameter values) rather than just around a fixed critical threshold,
as in isotropic percolation \cite{Roy2003,Pascual2005}.  Thereby, the
influential concept of ``robust criticality'' was introduced by
Pascual and colleagues to describe this phenomenon
\cite{Roy2003,Pascual2005}.  Other models with only two states
(occupied and empty) and some type of facilitation mechanism --as, for
instance, a positive feedback between vegetation and water supplies--
were claimed to support the idea of robust criticality
\cite{Scanlon2007,Manor-Shnerb-2008b}. From a theoretical viewpoint,
it is still not clear what is the origin of such generic or robust
scaling, nor if it is just a transient (not asymptotic) effect or if,
instead, it describes true scale invariance. Actually, van den Berg et
al. cast doubts on the theoretical foundations of the concept of
``robust criticality'' by analytically proving that in a simple
version of these models true scaling occurs only at a single point in
parameter space, i.e. there is an underlying sharp phase transition
and not a broad range of criticality \cite{van2011,van2015}. Thus,
understanding the origin of broad (not fine tuned) scaling in this
type of systems remains a challenge for theoreticians.

The aim of the present work is therefore to scrutinize whether generic
scaling --occurring in broad regions of the parameter space-- can
actually emerge in very simple dynamical models \cite{Henkel-book}, as
well as to help uncovering the basic mechanisms promoting the
emergence of scale invariance in vegetation patterns.  To this end, we
introduce a simple spatially-explicit individual-based model of
vegetation dynamics and analyze its behaviour. In particular, we
examine if the emergence of scale-free vegetation patches requires a
fine tuning of the control parameter (i.e. if it behaves as a standard
percolation-like transition) or if, on the contrary, scale-free size
distributions emerge in a robust way for a broad range of parameter
values.  We show that implementing temporal environmental variability
in the model --an ingredient inherent to actual ecological systems and
of special relevance in semiarid environments \cite{savanna}-- induces
the emergence of broadly distributed cluster sizes (that can be fitted
as power laws) in a wide region of the parameter space, while such a
region shrinks to a single point in the absence of temporal
variability. Hence, generic scaling in this case emerges as a sort of
intermittent or fluctuating percolation phenomenon, in which clusters
nearby a percolation threshold fluctuate in size owing to external
variability.  Remarkably, we show that our approach --even if with
some limitations inherent to its simplicity-- is able to reproduce the
overall dependence on vegetation density of some empirically observed
non-universal scaling exponents, thus contributing to shed light on
the ecological mechanisms behind such a phenomenology.

\section*{Model building}

In two-dimensional percolation on a square lattice the percolation
threshold is $p^{occ}_c\approx0.5927...$ and right at such a critical
point the cluster-size distribution follows a power law
\begin{equation}
  P(s)=C s^{-\tau}
  \label{tau}
\end{equation}
where $C$ is a normalization constant and the exponent takes the
exactly known value $\tau=187/91 \approx 2.05$; \neww{in finite
  systems this power law is cutoff by a rapidly decaying scaling
  function.  \cite{Stauffer-Aharony,Christensen-book}. } Importantly,
the value of $p_c^{occ}$ can change if another type of lattice of
local geometry is considered, but the exponent $\tau$ remains
unchanged, i.e. it is universal.

The individual-based model that we analyze is defined on a
two-dimensional square lattice (mimicking the surface of a semiarid
landscape) in which each lattice site, $j$, (i.e. each spatial
location) can be in one of two possible states: empty, $n(j)=0$, or
occupied $n(j)=1$.  At each discrete time step an
occupied site is randomly selected and, with probability $p$, a new
offspring is created at a random nearest-neighbor, provided it was
empty \cite{Marro,Henkel-book}.  Alternatively, with probability $1-p$
the selected occupied site is emptied (i.e. the individual dies).  The
time counter is increased as $t \rightarrow t + \Delta t$ with
$\Delta t =1/N(t)$, where $N(t)$ is the total number of occupied sites
at time $t$. This way, we ensure that in a (Monte Carlo) step all
sites are updated once, on average. We consider a system size
  of $N=1024\times 1024$, unless otherwise indicated.

  These stochastic rules define the ``contact process'', a
  prototypical model that captures the essence of a large variety of
  spreading systems in physics, biology, epidemiology,
  \cite{Marro,Henkel-book,GG} as well as in spatial ecology
  \cite{Durrett-Levin,Neuhauser,Durrett2009,Paula,Paula-Q,Bonachela-Munoz-Levin,JSP}.
  It can be summarized as the following set of {processes}:
\begin{equation}
  \arraycolsep=1.6pt\def\arraystretch{1}
 \begin{array}{cccc}
   A &\xrightarrow{p} & 2A & (birth), \\
 A   &\xrightarrow{1-p} & 0 & (death)
 \end{array}
 \label{test}
\end{equation}
where $A$  represents individual ``plants'' in our interpretation.
\begin{figure}
\centering
\includegraphics[width=0.75\textwidth]{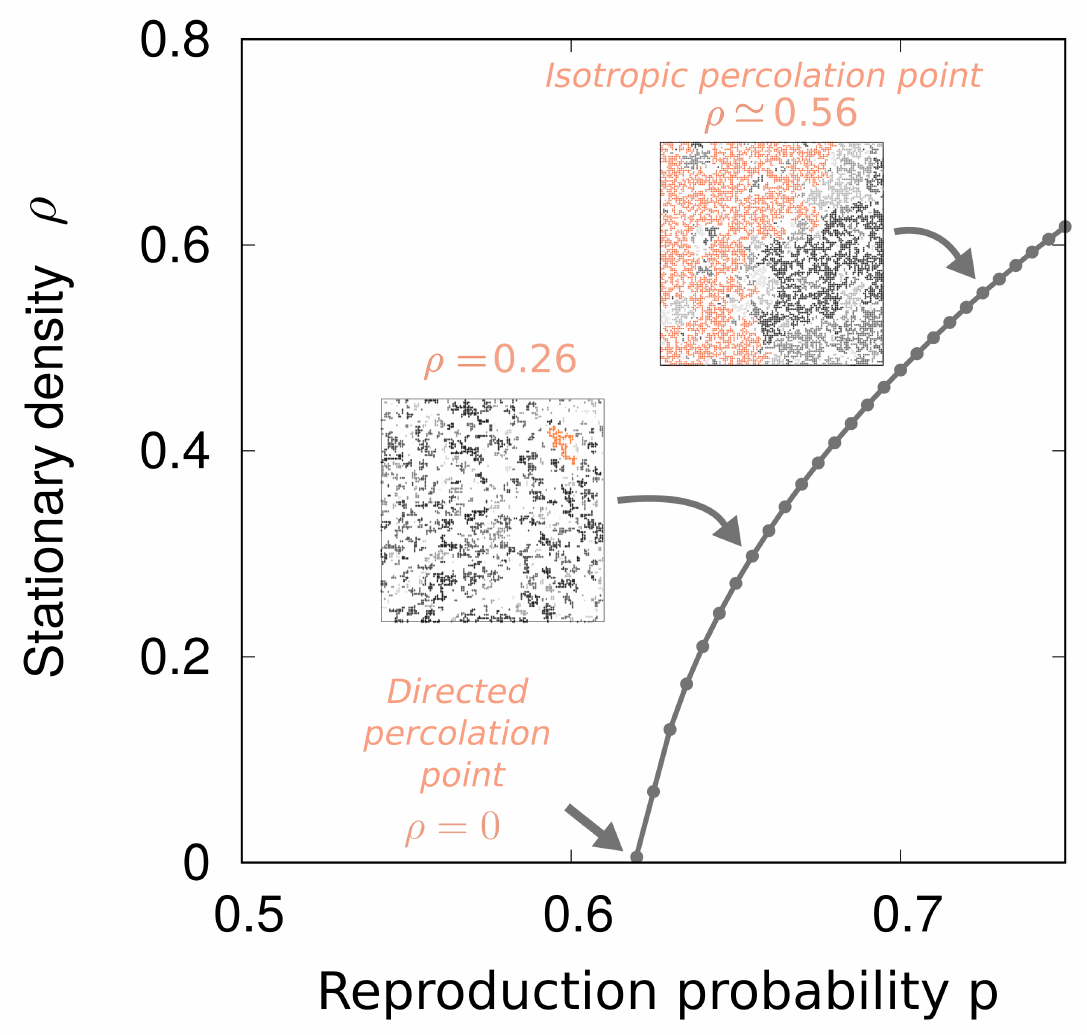}
\caption{{\textbf{Phase diagram of the standard or pure contact
      process}}.  At the directed-percolation critical point
  $p_c \approx 0.6225$ (separating empty from populated states), the
  stationary density is $\rho=0$, and it is not possible to measure
  any non-trivial steady-state spatial cluster distribution.  Observe
  that within the active phase --in which ``activity'' persists
  indefinitely across time-- clusters of adjacent active sites may:
  {\bf(i)} not percolate in space (as occurs, e.g. for $p=0.65$, which
  corresponds to $\rho\simeq0.26$) or {\bf(ii)} percolate in space,
  deeper into the active phase (as occurs for e.g. $p=0.725$,
  corresponding to $\rho\simeq0.56$). In the insets, we have
  highlighted (in orange) the largest cluster in two representative
  snapshots illustrating either the absence (for
  $\rho\simeq0.26$) or the existence of a percolating cluster (for
  $\rho\simeq 0.56$).}
\label{phase_diagram_ordered}
\end{figure}
In the special case of a well-mixed population, in which all sites are
taken to be neighbors of all others, and assuming an infinitely large
system (i.e. in the ``mean-field'' limit) the dynamics of the overall density of occupied sites $\rho(t)=\sum_{j=1}^{N} n(j)/N $ obeys
\begin{equation}
 \dot{\rho}(t)=p \rho(t)(1- \rho(t))- (1-p) \rho(t)= (2p-1) \rho(t)- p\rho^2(t) 
\label{MF}
\end{equation}
in unspecified time units \cite{Binney,Henkel-book,GG}.\footnote{This
  mean-field equation is straightforwardly derived by performing a
  size-expansion of the Master equation defining the model while
  keeping just the leading/deterministic term \cite{vanKampen}.}  This mean-field
equation exhibits a critical point at the reproduction probability
$p_c=1/2$ above which the population survives for arbitrarily long
time in sufficiently large lattices, reaching a stationary density
$\rho_s=(2p-1)/p$. Below such a value, the only stable solution is the
empty state $\rho_s=0$. Therefore, the critical point separates an
empty (or ``absorbing'') phase from an occupied (or ``active'') phase,
i.e. this is a \emph{directed-percolation} critical point
\cite{Henkel-book,Marro,Odor,GG}.

This type of exact solution cannot be straightforwardly extended to
analytically solve the case of a two-dimensional lattice in simple
terms; however, it is well-known that the overall phenomenology is
qualitatively (though not quantitatively) preserved
\cite{Henkel-book,Marro,Odor,GG}. This is illustrated in Figure
\ref{phase_diagram_ordered}, which shows the phase diagram for the
two-dimensional contact process resulting from computer
simulations. Observe, in particular, that the directed-percolation
critical point --separating the empty from the active phase-- is
shifted to $p_c \approx 0.6225 > 0.5$ due to fluctuations, i.e. to the
finite size of local neighborhoods.

Most of the literature on the contact process is focused on analyzing
the non-trivial features of this directed-percolation absorbing-active
phase transition \cite{Marro,Henkel-book,Odor,GG}.  However, here we
are interested in another ``hidden'' critical point whose very
existence is usually ignored. This, as sketched in
Figure 1, is a percolation-like transition above which a giant cluster of occupied
sites percolates in space \cite{van2011,van2015}.  More specifically,
this second transition is --as it will be shown below-- a
standard/isotropic \emph{percolation} phase transition above which, at
any given time within the steady state, the probability to find a
percolating cluster of adjacent occupied sites spanning the whole
two-dimensional space does not vanish
\cite{Stauffer-Aharony,Christensen-book}. In two-dimensions, one needs
to reach stationary activity densities of the order of
$\rho \approx 0.56$ (deep inside the active phase, i.e.
$p\approx 0.725 >p_c$) to observe the emergence of a spatially
percolating cluster (see Figure \ref{phase_diagram_ordered}). Let us
stress once more, that this transition at which clusters
\emph{percolate in space} is conceptually different from the
absorbing-to-active directed-percolation transition, at which clusters
of occupied sites can \emph{percolate in time}.

Hereafter, we study the nature of this (isotropic) percolation
transition emerging within the active phase of the standard, or
``pure'', contact process. However, our main focus will be on a
variant of the contact process implementing external/environmental
variability in the model, which is defined as follows: In the
\emph{temporally-disordered contact process}, at each Monte Carlo
timestep the reproduction probability, $p$, takes a
uniformly-distributed random value in the interval
$[\bar{p}+\sigma,\bar{p}-\sigma]$; thus, the parameter $p$ of the pure
contact process becomes a time-dependent function, $p(t)$\footnote{Let
  us note that one could have also defined the model by introducing
  temporal disorder in the ``microscopic'' transition rates in a
  continuous-time implementation of the contact process rather than in
  the probabilities \cite{Henkel-book}. We do not expect such a change
  to qualitatively affect the main results presented here.}.  The mean
value, $\bar{p}$ is the control parameter and is kept constant in each
simulation. In all cases, $\bar{p}$ is constrained to lie within the
interval $[\sigma,1-\sigma]$ so that $p(t) \in [0,1]$; on the other
hand, $\sigma$, characterizing the environmental variability, is
typically fixed (unless explicitly stated) to a value $\sigma=0.25$.

\section{The (isotropic) percolation transition in the standard
  contact process}
We start our analyses by focusing on the percolation transition of the
``pure'' contact process --without additional heterogeneity-- as it
will be used as a benchmark for further analyses.  The fundamental
difference from  the standard/isotropic percolation model is that the
contact process and its variants are dynamical models and the states
of nearby sites are correlated. Thus, the question we address now is
whether such a difference is a relevant one, i.e. whether the
universal aspects of the percolation transition observed within the
active phase of the contact process are equivalent to those of
isotropic percolation despite the existence of site-to-site
correlations in the first case.  In order to provide an answer to this question
we resort to both, computational analyses and analytical/heuristic arguments.

The first observation to be made is that in the pure model there is a
one-to-one correspondence between the control parameter $p$ and the
observed averaged density $\rho$ (see Figure 1 black curve), and as a
result they are interchangeable \footnote{This equivalence holds
  exactly for infinitely large system sizes, i.e. in the thermodynamic
  limit.}. Thus, we rely on the average density $\rho$ in the
forthcoming analyses, since it bears closer resemblance with what
empirically measured in field analyses.

\begin{figure}
\centering
 \includegraphics[width=\textwidth]{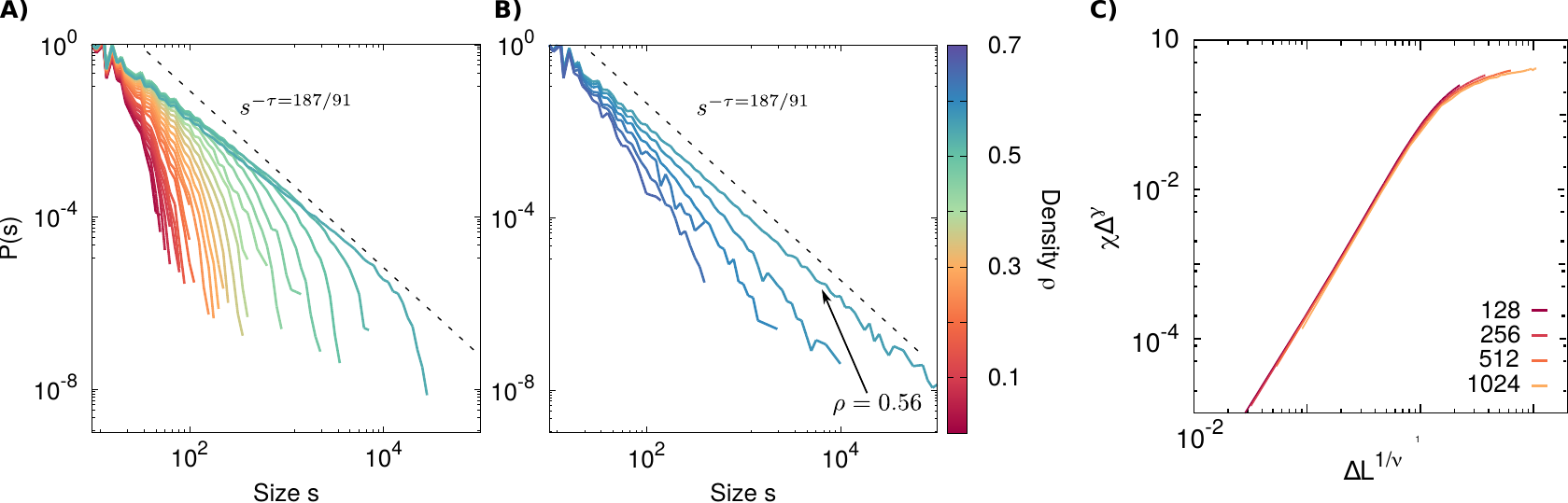}
 \caption{{\textbf{Standard isotropic percolation scaling within the
       active phase of the pure contact process ($\sigma=0$).}} The
   scaling in standard percolation at the critical occupation
   probability threshold $p^{occ}_c$ is characterized by a
   cluster-size distribution $P(s)$ such that $P(s) \propto s^{-\tau}$
   with critical exponents $\tau=187/91\approx 2.05$, and the scaling
   relationship
   $\chi=|p^{occ}-p^{occ}_c|^{-\gamma}\tilde{\cal{F}}(|p^{occ}-p^{occ}_c|
   L^{1/\nu})$ with critical exponents $\nu=4/3$ and $\gamma=43/18$
   \cite{Stauffer-Aharony,Christensen-book}. We show $P(s)$ for low
   ({\textbf{A}}) and high ({\textbf{B}}) densities (plotted
   separately for the sake of clarity) as obtained in computer
   simulations of the contact process on a square lattice, for
   different values of $\rho$ (corresponding to different values of
   $p$) in the range $\rho \in [0.02,0.70]$ in increments of $0.02$
   (color coded)). \neww{~Observe that strong deviations from  a
     power-law behavior  appear 
     as soon as the system is moved away for the critical point,
     especially from below.}
   Panel ({\textbf{C}}) shows the collapse curve for
   four different system sizes, using the scaling exponents of
   standard/isotropic percolation.}
  \label{CP_collapse-Ps}
\end{figure}

Figure \ref{CP_collapse-Ps}A and \ref{CP_collapse-Ps}B show the
cluster-size distribution $P(s)$ for a wide range of such occupation
densities $\rho$ in the pure contact process as measured in extensive
simulations. At the percolation threshold $p \approx 0.725$
--corresponding to the critical occupation-density value
$\rho_c\approx 0.56$-- and only at this value, the cluster-size
distribution decays as a power-law, i.e.  clusters of broadly
different sizes emerge.  Furthermore, the scaling behavior is fully
compatible with the expected universal behavior of standard/isotropic
percolation; indeed, the fitted value for the cluster-size
distribution exponent $\tau$ in the contact process at the percolation
threshold is in excellent agreement with the above-mentioned value
$\tau \approx 2.05$ of standard isotropic percolation
\cite{Stauffer-Aharony,Christensen-book}.
Moreover, other quantities usually employed in studies of
standard/isotropic percolation can also be determined at the
percolation transition of the pure contact process. In particular,
right at the percolation threshold, the mean cluster size $\chi$,
defined as
\begin{equation}
\chi=\sum_{i}s_{i}^2/n_{occ},
\end{equation}
--where $i$ runs over all clusters, $s_i$ is the corresponding cluster
size and $n_{occ}$ is the total number of occupied sites-- is expected
to obey the general scaling law
\begin{equation}
\chi=\Delta^{-\gamma}\tilde{\cal{F}}(\Delta L^{1/\nu}),
\end{equation}
where $\Delta=|p^{occ}-p^{occ}_c|$ is the distance to the critical
occupation probability $p^{occ}_c$, $L$ the system linear size, and
$\nu=4/3$ and $\gamma=43/18$ are exactly-known critical exponents for
the two-dimensional isotropic percolation class
\cite{Stauffer-Aharony,Christensen-book}. This implies that there
exists a \neww{scaling function}, o master curve,
$\tilde{\cal{F}}(\Delta L^{1/\nu})$ to which the data should collapse
for different system sizes $L$ and $\Delta$ values (if the right
values are used for the exponents $\nu$ and $\gamma$).  Figure
\ref{CP_collapse-Ps}C shows such a curve collapse for the pure contact
process at different values of $p$ --around the critical percolation
point $p \approx 0.725$ (corresponding to $\rho_c \approx 0.56$).
\new{Together, the three panels of Figure \ref{CP_collapse-Ps} show
  that the pure/standard contact process (i.e. with $\sigma=0$)
  displays a scaling behavior for a critical density
  $\rho_c \approx 0.56$ that is fully compatible with that of the
  standard/isotropic percolation class.}

In order to rationalize this conclusion from a theoretical viewpoint,
one can argue as follows. At the percolation critical point --which
lies well inside the active of the pure contact process-- occupied
sites are not randomly placed as they are generated by a dynamical
process that generates spatial correlations \cite{Henkel-book}.
However, as the dynamics at such a density is far away from the
absorbing-active phase transition (where long-range correlations are
well-known to emerge), spatial correlations are only short-ranged. As
a consequence --employing a renormalization-group perspective-- at
sufficiently ``coarse-grained'' scales, the effective density becomes
uncorrelated from one region to another. Thus, standard/isotropic
percolation scaling needs to emerge at sufficiently large scales. In
other words, the presence of weak short-range correlations is expected
to affect non-universal features such as the location of the critical
occupation density, but does not affect the large-scale universal
behavior of percolation, in perfect agreement with our computational
observations.

Summing up, the pure contact process exhibits scale invariance in its
cluster-size distribution if, and only if, it is fine tuned to have a
density close to the percolation threshold. Around such a density it
exhibits scaling features of standard percolation, and no trace of
generic scale invariance for arbitrary density values, for which 
exponential decay (with a characteristic scale) is observed.

\section{Percolation-like transition in the temporally heterogeneous model}

When temporal heterogeneity ($\sigma > 0$) is introduced in the model,
the scenario described in the previous section for the case of the
pure contact process ($\sigma=0$) is significantly altered, as we show
in what follows.  For the sake of simplicity and without loss of
generality, we present results only for $\sigma=0.25$, but
analyses with other values have also been performed to verify the
robustness of our conclusions.

\begin{figure}
\centering
\includegraphics[width=\textwidth]{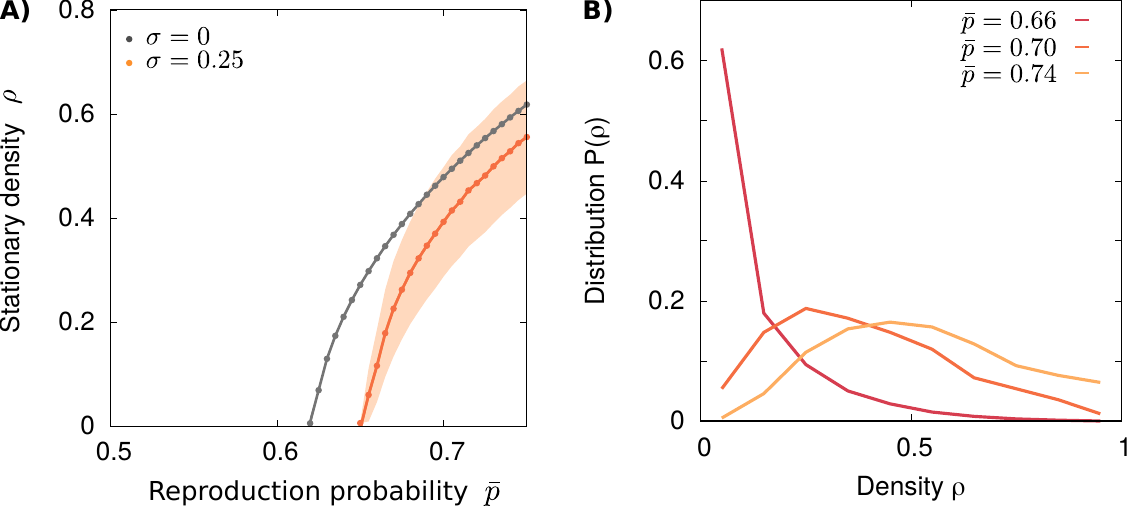}
\caption{ \new{\textbf{Phase diagram and density distributions of the
      temporally heterogeneous contact process}.  (\textbf{A}) Phase
    diagrams of the contact process in its ``pure'' ($\sigma=0$)
    version (plotted here, in light grey, for the sake of comparison)
    and the temporally heterogeneous ($\sigma=0.25$) version, in light
    orange.  For the temporally heterogeneous case, the displayed
    averaged density (points joined by a line) corresponds to the
    average over time as a function of the mean reproduction
    probability $\bar{p}$, with $\bar{p}\in[\sigma,1-\sigma]$, such
    that $p(t)=\bar{p}+ \sigma$ lay in $[0,1]$. In the heterogeneous
    case, the absorbing-to-active critical point is shifted to a value
    ($\bar{p}=p_c \approx 0.6550$ for $\sigma=0.25$) larger than that
    of the pure model (for larger values of $\sigma$ one observes
    larger shifts in the critical point; not shown). The plot also
    represents the standard deviation around the averaged density
    (orange shared area) for each $\bar{p}$.  (\textbf{B})
    Distribution of density values (histograms) as measured for for
    three different values of $\bar{p}$. These distributions remain
    broad even in the thermodynamic limit: i.e. there is no one-to-one
    relationship between $\bar{p}$ and the measured value of $\rho$, as this
   last fluctuates in time, even for infinitely-large systems.}}
\label{phase_diagram_disordered}
\end{figure}

\new{The first important observation to be made is that --opposite to
  the pure version of the model-- now there is not a one-to-one
  correspondence between the control parameter $\bar{p}$ and the
  resulting density $\rho$. Actually, as illustrated in Figure
  \ref{phase_diagram_disordered}, fixing $\bar{p}$ one observes a
  probability distribution of $\rho$ values that does not become
  sharp, even in the infinitely-large system-size limit \footnote{This
    property is shared by many other statistical physics models in
    which parameters change in time (see e.g. \cite{Alonso}).}. Owing
  to the intrinsic time variability of the control parameter, $p(t)$,
  the systems keeps on jumping between different values of $p$ --each
  one with a preferred stationary density-- and, thus, the overall
  density fluctuates --without relaxing to a precise fixed value-- even in
  the thermodynamic limit. This is also reflected in Figure
  \ref{phase_diagram_disordered}A in which we depict the resulting
  averaged density for each $\bar{p}$ (orange curve) together with its
  standard deviation (shaded region).}

As described e.g. in \cite{TGP}, the contact process with temporal
disorder exhibits also an absorbing-to-active phase transition. Here,
we find computationally that, for $\sigma=0.25$, the
absorbing-to-active critical point is located at
$\bar{p}_c\approx 0.6550$, a value slightly larger than the
corresponding value for the homogeneous (or ``pure'') case
$p_c\approx 0.6225$, as illustrated in
Figure \ref{phase_diagram_disordered}); \new{for larger values of
  $\sigma$ one can observe progressively larger shifts in the location
  of the critical point \cite{TGP} .}  Let us also mention, that this
temporally-disordered model exhibits an unusual scaling behavior,
including logarithmic behavior for some quantities (see
\cite{TGP,Vojta2015}).

As already explained, as soon as $\sigma > 0$ the distribution of
  possible values of the overall density $\rho$ is not a
  delta-function around a given density value (see Figure
  \ref{phase_diagram_disordered}B).  A straightforward consequence of
  this is that the ``constant-$\bar{p}$ ensemble'' and
  ``$constant-\rho$ ensemble'' are not equivalent: it is not the same
  thing to collect statistics of cluster sizes for a fixed value of
  $\bar{p}$ than to do so for a fixed value of $\rho$. This is in
  stark contrast with the standard pure case in which both ensembles
  are equivalent, as in the thermodynamic limit the distribution of
  $\rho$ possible values becomes a delta function for any value of
  $\bar{p}$ \cite{equivalence,Henkel-book,TGP}.

\begin{figure}
\centering
 \includegraphics[width=\textwidth]{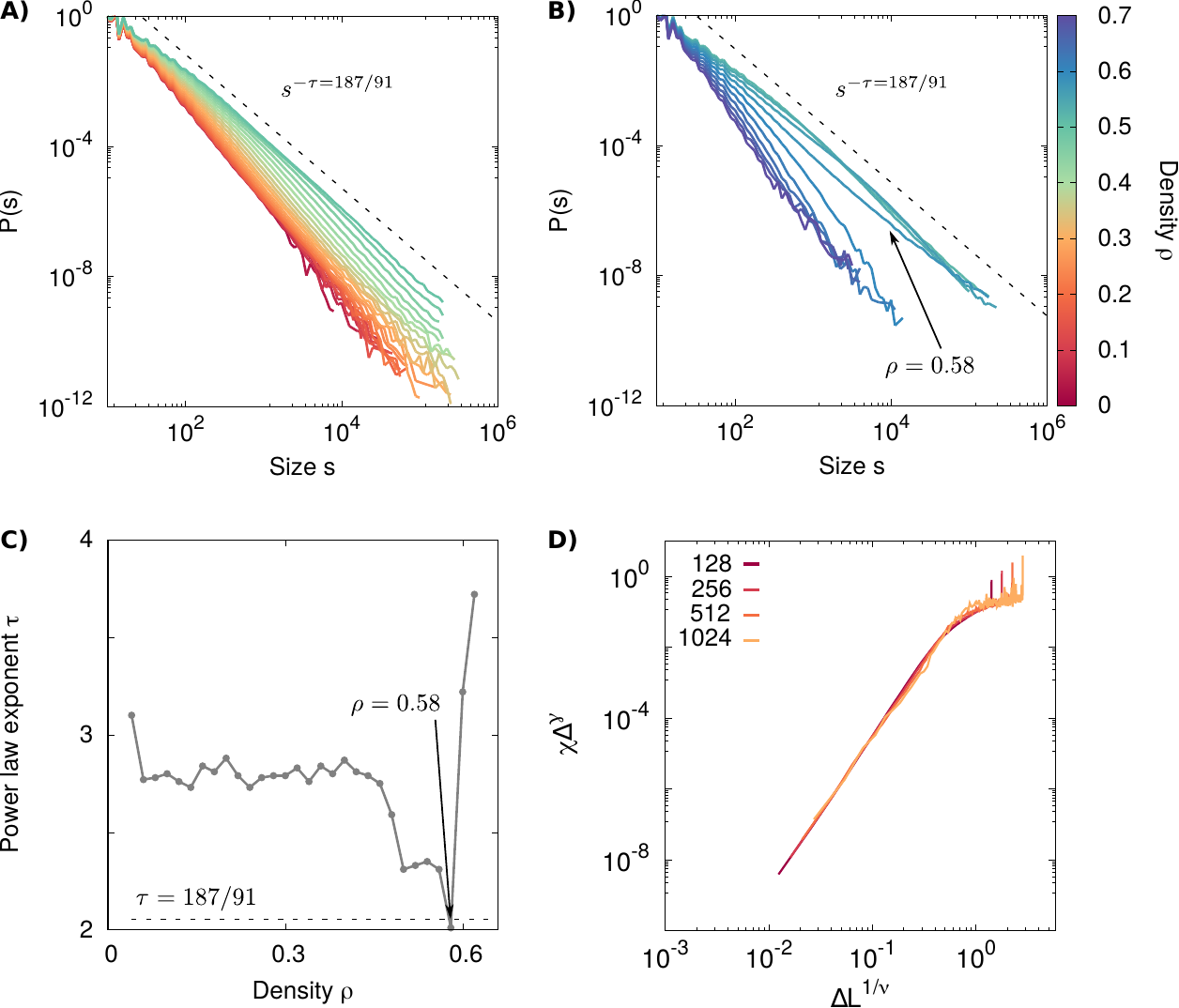}
 \caption{{\textbf{Percolation within the active phase of the
       temporally heterogeneous contact process.}}  \neww{ As in
     Figure \ref{CP_collapse-Ps}, we represent cluster-size
     distributions, for low density values $\rho \in [0.02,0.5]$
     ({\textbf{A}}) and for high-density values $\rho \in [0.52,0.7]$
     ({\textbf{B}}) (color coded).  Observe that, remarkably --on the
     contrary to the pure contact process case-- the curves do not
     bend down quickly as soon as the system departs from the critical
     point; actually, the curves exhibit heavy tails and power-law
     functions can be fitted for a wide range of sizes (at least two
     or more decades for all considered $\rho$ values, and up to four
     decades around the critical point).  (\textbf{C}) Exponent value
     of the power-law fit, $\tau$, as a function of $\rho$, exhibiting
     a non-monotonic dependence. The exponent $\tau$ starts to
     increase at $\rho=0.58$ with a value $\tau \simeq 2.01$, slightly
     smaller than the expected for the standard percolation
     $\tau=187/91 \simeq 2.05$; this ``noisy'' curve suggests
     that there are significant errors in the determination of $\tau$
     exponents. (\textbf{D}) Considering the value $\rho=0.58$ as an
     estimation of the critical percolation point, we obtain a rather
     good scaling collapse for $\chi=\Delta^{-\gamma}\tilde{\chi}(\Delta L^{1/\nu})$ with
     exponent values $\nu\simeq3$ and $\gamma\simeq4$, much larger
     than those of standard/isotropic percolation $\nu=4/3$ and
     $\gamma=43/18$, revealing a different type of scaling.}}
  \label{TCP_collapse-Ps}
\end{figure}

In order to properly compare the computational results of the
temporally-disordered model with field observations of vegetation
cover --for which the only empirically measured quantity is the
corresponding vegetation density $\rho$ \cite{Scanlon2007,Kefi2007}--
it is more pertinent to consider a``$constant-\rho$'' ensemble.  In
practical terms, to obtain results for a given density $\rho$, we run
computer simulations for different (homogeneously sampled) values of
$\bar{p}$ and for sufficiently long times (typically $t>10^3$) as to
reach a statistical steady state. Using such timeseries, we take
``snapshots'' of different configurations across time (taking them
sufficiently separated in time as to be uncorrelated) and collect the
statistics of patches sizes \emph{conditioned} to the value of the
overall density $\rho$ ($\pm0.0005$). Notice also that if an infinite
(system-spaning) cluster exists, it is not considered for the statistics
of \emph{finite} clusters \cite{Christensen-book}.

\begin{figure}
\centering
\includegraphics[width=\textwidth]{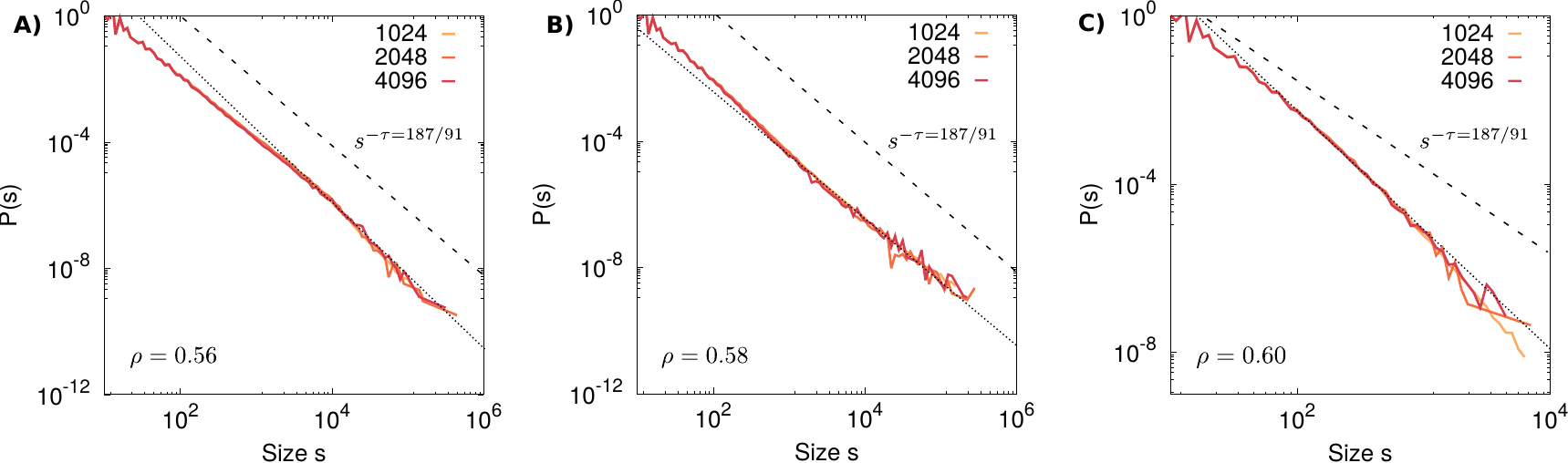}
\caption{\neww{{\textbf{Analyses of cluster-size distributions for
        different system sizes.}  The plots show the computationally
      measured cluster-size distributions $P(s)$ for three different
      density values ({\textbf{A}}) $\rho=0.56$, ({\textbf{B}})
      $\rho=0.58$, and ({\textbf{C}}) $\rho=0.60$. For each density,
      distributions are plotted for three different system sizes
      $N=1024, 2048$, and $4096$, respectively. Power laws for the
      standard percolation $\tau={-187/91}$ (dashed lines) and fitted 
      (dotted lines) exponents are shown for illustration. These
      results prove the robustness of the values for the exponent
      $\tau$ reported in Figure 4.  }} }
  \label{FSS}
\end{figure}
In what follows, we present the results of computational analyses for
the resulting $ P(s)$ distributions for different values of
$\rho$. \neww{Let us underline before going forward that, this way of
  proceeding considers configurations (``snapshots'') generated under
  different dynamical conditions to generate one single probability
  distribution function for each density value.}  In other words, the
obtained results for a fixed value of $\rho$ correspond to a mixture
of different dynamical processes, e.g. with different values of
$\bar{p}$ and with diverse realizations of the temporal disorder, but
nevertheless, well-defined probability-distribution functions emerge.

Using the statistics collected this way, one can find that the best fit to $P(s)$ for
each value of $\rho$.  Given the subtleties usually encountered when
determining whether a given distribution follows a power law, let us
stress that in the forthcoming analyses we employ the standard
most-stringent statistical tests (following the methods introduced in
\cite{Clauset}; see Appendix) to fit power laws and to decide
whether they constitute a better fit than e.g. exponential
distributions.  Careful analyses reveal that --for a large variety of
values of $\rho$-- cluster-size distributions conform to power-law
curves for at least two (and up to four) decades in size (see Figure
\ref{TCP_collapse-Ps}A and B, and Table \ref{table:fit_averaged}). In
particular, one can estimate the parameters $C, \tau$, $s_{min}$, and
$s_{max}$ in Eq.(\ref{tau}) that maximize the normalized
log-likelihood of the fit (see Appendix).

\neww{We observe the presence of fat tails in the distributions $P(s)$
  for different densities (observe also, that there are crossovers,
  clearly visible e.g. for $\rho=0.58$). The tails of these
  distributions can be fitted as power-law functions with continuously
  varying exponents for a wide range of $\rho$ values.  Indeed, eye
  inspection of Figure \ref{TCP_collapse-Ps}A and B already suggests
  that the exponent $\tau$, characterizing the power-law decay,
  displays a non-monotonic variation with population density $\rho$
  (see also Figure \ref{TCP_collapse-Ps}C).  The fitted value
  decreases from $\tau\simeq 3.02$ (for $\rho= 0.02$) down to
  $\tau\simeq 2.01$ (for $\rho \approx 0.58$) and increases again up
  to $\tau\simeq3.53$ (for $\rho \approx 0.7$) (see Figure
  \ref{TCP_collapse-Ps}C and Table \ref{table:fit_averaged}). }

Moreover, in order to verify the robustness of these results, in
Figure \ref{FSS} we present results for increasing system sizes,
confirming that the measured power-law fits are not finite-size
dependent, i.e. they are robust. Notice, in all cases, the lack of
strong downward bends even if the considered system sizes are not
very large; i.e. the distributions are consistently fat tailed.

Thus, remarkably, opposed to the case of the pure contact process--
power-laws fit well the cluster-size distributions $P(s)$ not only at
a unique (critical) point but also in a wide parameter range above and
below such a value, i.e.  we observe scale-free cluster sizes through
a wide range of densities, \neww{at least, up to numerical
  resolution}. Owing to this peculiarity, the percolation threshold
$\rho_c$ cannot be easily obtained --as usually done in numerical
analyses-- just by looking for the precise value of the control
parameter for which scale-free cluster-size distributions emerge
\cite{TGP}.  Therefore, in order to obtain an estimate of the critical
point it is necessary to employ some criterion; in particular, we
define the critical percolation point as given by the value of $\rho$
for which \neww{$\tau$ changes its tendency, from decreasing to
  increasing, i.e.  $\rho \approx 0.58$ (see Table I for further
  details on the fits). We need to emphasize that this is only an
  approximation; more precise estimations of the percolation critical
  point are not easy  to obtain, not even employing finite-size scaling analyses
  \cite{TGP,Vojta2015}.}

The cluster-size distribution exponent fitted at \neww{$\rho=0.58$} is
\neww{$\tau\simeq 2.01$, slightly smaller} than the expected value in
two-dimensional percolation $\tau\approx 2.05 $ (see Table
\ref{table:fit_averaged}).  Moreover, for the present estimation of
the critical point, we obtain a scaling collapse (see Figure
\ref{TCP_collapse-Ps}D) similar to that of the homogeneous case,
represented in Figure\ref{CP_collapse-Ps}C but with exponents
\neww{$\nu\simeq3$} and $\gamma\simeq4$ that are considerably larger
than the percolation values ($\nu=4/3\approx 1.33...$ and
$\gamma=43/18\approx 2.39...$), suggesting a different type of
critical behavior (or universality class). However, it is important to
remark that by slightly shifting the critical point estimation,
e.g. \neww{to $\rho=0.56$, for which $\tau \approx2.31$}, we obtain a
similar-quality curve collapse with slightly different exponents
(\neww{$\nu\simeq3.5$} and $\gamma\simeq4$).  This suggests the
existence of relatively large error bars in the numerically-estimated
critical exponents.

  \neww{ Thus, at the light of all these difficulties, it is clear
    that a more precise and careful determination of the critical
    point, the associated critical exponents, and their possible
    scaling relationships is still needed to obtain a deeper
    understanding of this novel type of percolation phase transition.
    This is, however, beyond the scope of the present manuscript and
    is left as a challenge for future work.  }

\section{Individual-snapshot properties of the temporally
  heterogeneous model}

We have shown that the contact process equipped with temporal disorder
exhibits scale-free cluster-size distributions without the need of
careful fine tuning to a threshold point. Nevertheless, in order to be
able to make a more direct comparison with empirical observations, an
important caveat remains. The results obtained so far correspond to
\emph{ensemble averages}, where many different realizations of the
system --e.g. at different times-- are considered together with the
only restriction of a fixed $\rho$ value. However, empirically
measured exponents of vegetation cluster-size distributions are
determined for individual ``snapshots''.
\begin{figure}
\centering
\includegraphics[width=\textwidth]{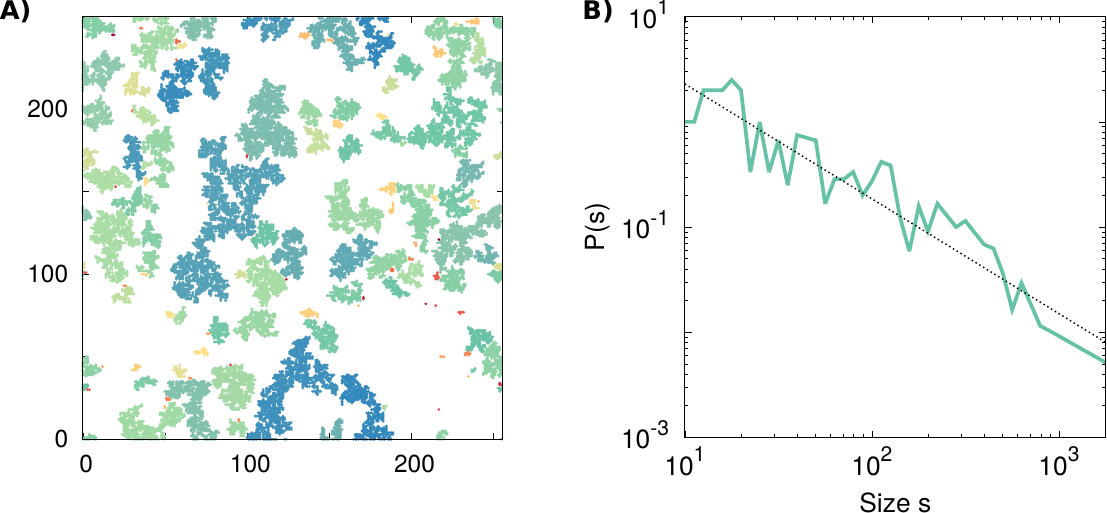}
\caption{\textbf{Clusters and cluster-size distribution in the
    temporally heterogeneous model for a density $\rho=0.3$ far from
    the percolation threshold.}  ({\textbf{A}}) Snapshot of
  the system (with size $256\times256$) at some given time at which
  the density takes the value $\rho=0.3$; the color code indicate
  cluster size: from red for the smallest to blue for the
  largest ones. (\textbf{B}) Cluster-size
  distribution (histogram) as obtained from panel (\textbf{A}); the distribution
  is obviously noisy given the relatively small system size, but it is
  clearly a broad one, and can be well fitted as a power law (black dotted
  line).}
  \label{Snapshot1}
\end{figure}

To bridge this gap, we computed the statistics of individual or
``instantaneous'' snapshots within our model.  In other words, we
performed a set of simulations of the model for diverse (homogeneously
sampled) values of $\bar{p}$ and took instantaneous pictures of the
system configuration at the steady state. For each of such snapshot,
$j$, we computed its overall density and measured its associated
cluster-size distribution (histogram), $P_j(s)$. \neww{Obviously,
  these distributions --not being ensemble averaged-- are much noisier
  than their ensemble-averaged counterparts.}  However, simple eye
inspection of Figure \ref{Snapshot1}B already reveals that the
cluster-size distribution of an instance, for an individual snapshot,
can span over a large range of cluster sizes, and can, in fact, be
fitted by a power-law, even for a density far away from the percolation
threshold (such as $\rho=0.3$ in this example).

\begin{figure}
  \centering
  \includegraphics[width=\textwidth]{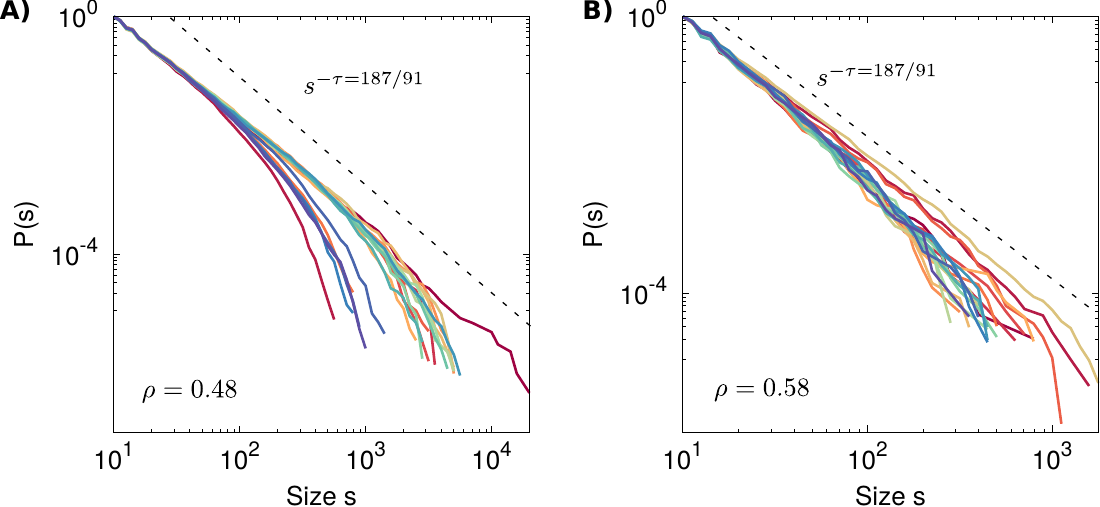}
  \caption{{\textbf{Cluster-size distributions for individual
        snapshots of the temporally heterogeneous model.}}
    Cluster-size distributions are shown  for a few
    snapshots with diverse density values: ({\textbf{A}})
    \neww{$\rho=0.48$, and ({\textbf{B}}) $\rho=0.58$,} respectively.
    Due to the high variability of cluster size distributions,
    individual realizations --rather than ensemble averages-- are more
    adequate to characterize snapshots of real systems. Observe
      that approximate power-law distributions appear for snapshots of
      different densities but also, curves deviating from
      power-laws are seen, especially for $\rho=0.48$.}
  \label{Snapshots2}
\end{figure}

A collection of individual-snapshot cluster-size distributions are
shown in Figure \ref{Snapshots2} \neww{for two different densities:
  $\rho=0.48$ and $\rho=0.58$}.  This figure illustrates that even for
a fixed density $\rho$ there is a high variability in the shape of the
histograms of cluster-sizes and that single snapshots may differ
notably from the ensemble-averaged (constant-$\rho$) distribution. In
particular, fast exponential-like and slow power-law-like decays can
be identified for individual snapshots. Remarkably, even if
exponentially decaying curves do exist, especially for low densities,
power-laws are present in snapshots for most densities.  These
power-law fits are more prominent and extend for a broader
size-interval in the neighborhood of the percolation transition (see
Table II for more details on the fits).

We also compared the quality of fits to power-law functions with those
to e.g. exponential-distribution $P(s)=C e^{-\alpha s}$.  For many
snapshots --especially when close to the percolation critical point--
power-law distributions constitute a better fit to the data than
exponential ones (i.e. power-laws give a higher normalized
log-likelihood) but this is not always the case, especially for low
densities $\rho\in[0.04,0.24]$ (see fourth column of Table
\ref{table:fit_snapshots} and Appendix A for further
details). Nevertheless, even for densities in this interval, power-law
fits are better in at least the $60\%$ of the cases (see Table
\ref{table:fit_snapshots} and Appendix).  On the other hand, for
intermediate densities, power-law fits outperform exponential ones in
almost each one of the snapshots. Observe that in spite of the existing
limitations  (i.e. the finite system size and the lack of an
averaging procedure), power-laws are observed for at least $1.5$ decades
in logarithmic scale in most of the cases (fifth column of Table
\ref{table:fit_snapshots}).

\neww{To avoid confusion, let us underline that some of these
  effective exponents are smaller than $2$ and that power laws with
  exponents smaller than $2$ do not have a well-defined averaged value
  when integrated to arbitrarily large values of the variable.  This
  means that the fits are just approximated ones and cannot possibly
  extend to arbitrarily large cluster sizes (see \cite{tau} for an illuminating 
  discussion of this and related issues).}
 
\begin{figure}
  \centering \includegraphics[width=0.7\textwidth]{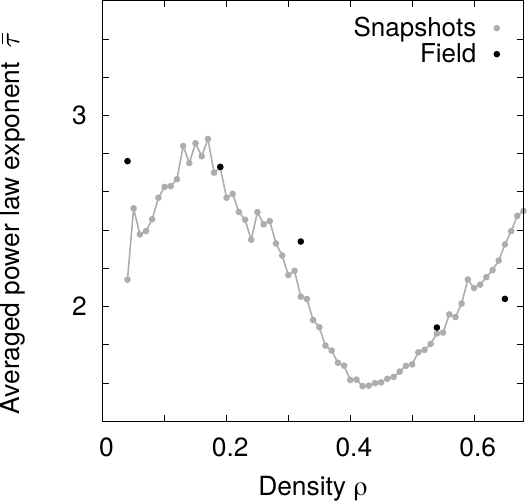}
  \caption{{\neww{\textbf{Averaged individual-snapshot cluster-size
          distribution exponents are in reasonable agreement with the
          ones observed in field measurements.}  We fitted power-law
        regimes to cluster-size distributions $P(s)$ for $200$
        different snapshots at different densities $\rho$ for density
        values $[0.04,0.7]$ equally spaced in increments of size
        $0.01$ (see Appendix and Table \ref{table:fit_snapshots} for
        more details).  The plot shows the resulting averaged exponent
        value $\overline{\tau}$ (grey line) of snapshots best
        described by power-law cluster-size distributions
        (distributions with exponential tails are discarded) as well
        as empirical data from field measurements
        \cite{Scanlon2007,Kefi2007} (black dots); these are in
        reasonably-good agreement with the model predictions (at
        least, for the relatively large value of the environmental
        variability, $\sigma=0.25$, considered here).}}}
  \label{Empirical}
\end{figure}

The average of the exponents obtained by fitting power-law
distributions for individual snapshots --obtained excluding samples
that are best fitted by exponentials-- exhibits a non-monotonic
behavior as a function of the vegetation density (see gray line of
Figure \ref{Empirical}) similar to that of the average in Figure
\ref{TCP_collapse-Ps}C.

\neww{It is important to emphasize that this effective averaged value
  of $\overline{\tau}(\rho)$ is not the same as the ensemble averaged
  value of $\tau(\rho)$ obtained in the previous section: here, we fit
  individual snapshots, discard those that comply better with
  exponential fits, and average the resulting set of $\tau$ values to
  obtain a final averaged effective exponent $\overline{\tau}(\rho)$;
  this is different from considering the averaged distribution
  including \emph{all} snapshots with a given $\rho$, and fitting
  a single exponent $\tau(\rho)$ to such a final distribution.
  Observe that the mean exponent $\overline{\tau}(\rho)$ reaches a
  minimum value $\overline{\tau} \simeq 1.58$ at a density of
  $\rho\simeq0.42$, below the percolation threshold, and increases up
  to $\overline{\tau} \simeq 2.8$ for larger and smaller
  densities\footnote{Note that, as discussed above, exponent values
    smaller than $2$ can appear.}.}

We also performed simulations for
    smaller values of the noise amplitude $\sigma$, for which the
    shape of the curves remain similar; in the limit in which $\sigma$
    goes to $0$ one obtains typically exponential fits, and one should
    recover power-law scaling only in a vanishing region around the
    critical point.

    Finally, we can compare the obtained averaged exponents
    $\overline{\tau}$ with available empirical data
    \cite{Scanlon2007,Kefi2007}.  In particular, the black dots of
    Figure \ref{Empirical} correspond to empirical data for different
    vegetation-cover densities. Remarkably, field data are roughly
    compatible with the model results (for $\sigma=0.25$) within one
    standard deviation (p-value $p=0.04<0.05$ of Pearson's $\chi^2$
    statistics). Of course, this can be just a coincidence, as we have
    no information on what the empirical value of $\sigma$ would be
    for the different actual data, nor it needs to be same for all of
    them. Still, we find it noteworthy that the overall dependency on
    vegetation density is reproduced by the present simple model.

\section*{Discussion} 

The study of vegetation patterns has a long tradition in ecology and
environmental sciences. This interest has been reinforced in recent
years owing to the possibility that the statistics of such
patterns can be used as a predictor of early warning of ecological
transitions or regime shifts, such as desertification in semiarid
environments. From the empirical side, two very influential papers
reported a decade ago that vegetation patterns in some arid and
semiarid ecosystems are scale free, thus lacking characteristic scales
\cite{Scanlon2007,Kefi2007}.  Puzzled by these field observations,
theoreticians were challenged to understand the origin of such
scale-free patterns and, as a result, diverse mechanisms have been
proposed to account for them. Different models put the emphasis on
existing plant-plant interactions (competition and facilitation),
feedback loops between vegetation dynamics and abiotic resources such
as water \cite{Scanlon2007,Manor-Shnerb-2008a}, or the presence of
spatial and/or temporal heterogeneity in the environment
\cite{Kefi2007,Shnerb2003-glass,Manor-Shnerb-2008b,Manor-Shnerb-2009}.

For the sake of generality, let us stress here that the theoretical
challenge of modelling approaches is to explain how scale-free
patterns can possibly emerge in a robust way, i.e. without the need of
invoking a high degree of parameter fine tuning. In particular,
experience in statistical physics tells us that scale-free patterns
typically emerge at very special points of the parameter space, such
as critical points. To circumvent this theoretical conundrum and in
order to explain generic power-law decays, two different types of
approaches are considered.  The first one consists in considering
mechanisms of self-organization, inspired in self-organized
criticality \cite{BTW,Pruessner-book,BJP,JABO1}, which would justify
why critical-like scale-free features appear without the apparent need
of parameter fine tuning (although they usually require an unrealistic
separation of time scales). This type of approach leads, however, to
universal --as opposed to continuously varying-- exponents.  The
second strategy consists in devising models in which scale-free
patterns --with their concomitant power law distributions-- emerge for
broad ranges of parameter values \cite{GG-generic}.  Such type of
models already exist in statistical mechanics (for an in-depth
discussion of these issues we refer to \cite{GG-generic}), but, to the
best of our knowledge, they are not well-suited to be translated to
the ecological problem under study here.

A particularly relevant idea in this context is that of ``robust
criticality'', stating that scale-free clusters of vegetation are akin
to percolation clusters at a percolation phase transition, but that
--for reasons to be fully clarified-- it could be observed in simple
models of vegetation dynamics without the need of fine tuning exactly
to the percolation transition
\cite{Pascual2002simple,Pascual2002cluster,Roy2003,Pascual2005,Pascual2006}.
In other words, it is claimed that these models exhibit a broad region
of power-laws around a transition point. However, the origin of such a
broad region of criticality is still controversial in the sense that
there is no actual theoretical understanding of why generic scale
invariance could possibly emerge.

In this paper we have proposed a very simple model that is able to
justify the emergence of \neww{very broadly distributed vegetation
  patterns}-- whose tails can be fitted as power laws, i.e. the
hallmark of scale invariance-- for a very wide range of parameter
values.  Our model is individual-based and spatially explicit, based
on the two-dimensional contact process, an extensively studied model
that is the simplest to mimic vegetation dynamics (as a
birth-and-death process). The contact process is well-known to exhibit
a quiescent-to-active phase transition, separating a quiescent phase
on which all activity/vegetation ceases/disappear, from another in
which activity/vegetation survives indefinitely with a given average
density. In other words, the contact process exhibits a
``directed-percolation'' type of phase transition separating phases
where activity may, or may not, percolate in time.

On the other hand, it is often neglected that there is another phase
transition hidden in this prototypical model. It occurs within the
active phase (as illustrated in Figure \ref{phase_diagram_ordered} and
\ref{CP_collapse-Ps}) and separates a phase in which the largest
cluster of active sites percolates in space from another one in which
it does not. We have confirmed computationally that, indeed, as
theoretically predicted by van den Berg \emph{et al.}
\cite{van2011,van2015} the transition is sharp, i.e. there is not such
a thing as a broad range of scaling for such a simple model.
Moreover, we have shown for the first time that such a transition is
perfectly described by the standard scaling theory of (isotropic)
percolation, with its well-known set of scaling exponents (see Figure
\ref{CP_collapse-Ps}). Let us emphasize that such a result is not
trivial as in standard percolation sites are occupied in a completely
random way, while in the active phase of the contact process activity
occupied sites are spatially correlated. Readers familiar with
renormalization-group theory could have correctly anticipated though
--employing renormalization-group reasonings-- that such short-ranged
correlations should not be able to affect the scaling exponents. Thus,
summing up, the standard contact process only exhibit scale-free
clusters at a very precise point in parameter space and is therefore
\emph{not} an adequate model to explain the broad/generic emergence of
scale-free vegetation patterns in a robust way.

Inspired by previous work of our group and others showing that
temporal variability can lead to generic power laws in some simple
models \cite{TGP,Vojta2015,TGP-Shnerb}, we moved on to study a variant
of the contact process implementing temporal variability in external
conditions, i.e. a randomly changing control parameter. With this
simple --and ecologically well justified-- novel ingredient the model
changes dramatically.

{\bf{ i) }} First of all, the percolation phase transition within the
active phase is not in the standard/isotropic percolation universality
class, i.e. its scaling exponents are significantly different form the
standard ones (see below for a further discussion of this), even if
further analyses are required to establish the precise values of the
exponents and the associated scaling relationships.

{\bf{ ii) }} Second, generic scale-free distributed clusters emerge on
average for a broad range of overall densities and not just at the
critical one. \neww{As in the previous point, further work is needed to
precisely establish the values of their asymptotic exponent values
and their dependence on $\rho$.}

{\bf{iii) }} Third, individual snaphots of the system state
--i.e. single configurations in which no averaging is performed-- can
also exhibit clusters of highly variable sizes that can be fitted as
power laws for a broad range of possible vegetation densities.  Even
more remarkably, the obtained power-law exponents are density
dependent --in agreement with what observed in empirical findings--
and exhibit --roughly speaking-- the same type of density dependence
as those observed empirically.

Let us now briefly discuss each of these three results.

{\bf{i) }} To understand in a qualitative way why the scaling at the
percolation point does not coincide with that of standard/isotropic
percolation, one could argue as follows. When unfavorable external
conditions appear, small clusters are very likely to be removed, while
large ones, although also affected by poor conditions, can remain
connected or divided in smaller-sized clusters.  When favorable
periods return, the latter clusters become more compact and/or are
re-merged with others. Contrarily, the removed clusters are not able
to re-appear from the empty sites, so that empty regions persist. As a
result of these effects, active sites in the active phase are
strongly correlated and not equivalent to a standard/isotropic
percolation process. Of course, in order to turn these hand-waving
arguments into a more formal proof, one would need to perform some
type of renormalization-group calculation, which is beyond our scope
here. However, let us remark, that usually the presence of temporal
disorder dramatically affects the universal behavior of well known
universality classes \cite{Vojta2015,Alonso} (including the
quiescent-active critical point of the contact process, which is
dramatically affected by temporal noise \cite{Moreira, Cafiero,TGP}).
Thus, it does not come as a surprise that our model equipped with
temporal disorder differs essentially from standard percolation. In
any case, a systematic account of this (novel?) type of scaling,
establishing all critical exponents and scaling relationships with
good accuracy and precision --and determining whether this really
constitutes a robust universality class-- is left for future work.

{\bf{ii) }} The emergence of a broad region of densities for which
scale-free behaviour emerges is in line with the idea of generic
scaling emerging in the presence of temporal variability. Actually, it
was recently shown that in the presence of temporal heterogeneity a
novel type of phase may emerge in systems with an absorbing-active
phase transition such as the contact process. Such a novel phase was
named \emph{ temporal Griffiths phase} \cite{TGP} in analogy with the
well-known \emph{Griffiths phases} that emerge in the presence of
spatial (quenched) heterogeneity \cite{Vojta}. In particular, it was
shown in \cite{TGP} (see also \cite{Savanna,TGP-Shnerb}) that for
models similar to the contact process studied here, some quantities
scale as power laws, not just at a critical transition point, but in a
whole broad region in parameter space in the presence of temporal
disorder, and exhibit an ``exotic'' type of scaling \cite{Vojta2015}).

{\bf{iii) }} Directly related to our findings is the important work of
Manor and Shnerb who discussed --in a mean-field setup different from
our spatially explicit one-- how multiplicative (environmental) noise
can lead to scale-free distributed cluster sizes
\cite{Manor-Shnerb-2008a,Manor-Shnerb-2009}. In a nutshell, each
existing cluster experiences a multiplicative random-walk process in
which the change in cluster size either increases or decreases at each
time step, but, crucially, it does so in an amount which is
proportional to its present size, i.e. in a multiplicative way. It is
very well-known that multiplicative processes lead generically to
power-law distributions (of sizes in this case)
\cite{Mitzenmacher,Solomon,Sornette-multiplicative}.
We believe that a multiplicative process similar to this one is at the
basis of the power-laws found in our model, but it can not be 
analyzed in simple mean-field terms.

Thus, we believe that the reason why our model produces generic
power-law distributions (at least in an approximate way) is twofold:
(i) on the one hand, nearby the standard percolation phase transition
clusters of rather different sizes are generated and (ii) owing to the
multiplicative process induced by changing environmental conditions
such clusters fluctuate wildly. The combination of these two features
leads to --at least approximate-- scale-free cluster distributions. In
other words, the model can exhibit a sort of \emph{intermittent
  percolation} in which the overall density fluctuates around the
critical (isotropic) percolation threshold. This sweeping through the
transition (or even near the transition) may suffice to generate
power-laws in a rather robust fashion \cite{Sornette1994}.  \neww{In
  any case, further work is still needed to confirm in a more clean
  cut way that the power laws observed here within a broad range of
  parameter values are indeed asymptotic ``true'' ones, observable at
  arbitrarily large scales; deeper theoretical understanding is still
  needed to advance in this direction.}

\section*{Conclusion} 

The conclusion of the present study is that a very simple
individual-based model with fluctuating external conditions is able to
reproduce in a simple but robust way the emergence of scale-free
vegetation patterns. The exponents associated with the concomitant
power-law distributions are non-universal and density dependent, in
agreement with what empirically observed.

From the theoretical-ecology perspective, this conclusion does not
imply that other mechanisms, such as plant-plant interactions or
plant-water feedbacks are not important to explain vegetation patterns
in general, but confirms that temporal variability is a key player to
explain the emergence of scale-free vegetation patterns in semiarid
environments.

\new{From a theoretical-physics perspective, a new type of percolation
transition has been uncovered here. It is our hope that this works
stimulates further analyses of its scaling properties and helps to
develop new applications to real-world phenomena.}

\section*{Appendix}

\subsection*{Fits of cluster-size probability distributions}

We fit cluster-size distributions $P(s)$ --both averaged over
different realizations and times and for individual snapshots-- as
power-laws $P(s)=C s^{\tau}$ within an optimal interval
$[s_{\min},s_{\max}]$. \neww{For each combination of the set of
  parameters $\tau, C, s_{min}, s_{max}$, we obtained the normalized
  log-likelihood assuming Poissonian counts
  $\ln L=(1/N)\sum_i[s_i \ln P(s_i)-P(s_i)+\ln(s_i!)]$, where $N$ is
  the number of non-zero cluster sizes in a given range of $s$
  \cite{Clauset,Klaus2011statistical,Alstott2014powerlaw}. We then
  selected the parameter sets that maximizes the normalized
  log-likelihood in each case. To avoid overfitting, we discard fits
  whose optimal range $[s_{\min},s_{\max}]$ includes less than a fixed
  number of points: for averaged distributions $P(s)$ (Figure
  \ref{TCP_collapse-Ps}) this range is fixed to $25$ points, while for
  snapshots (figure \ref{Snapshots2}) we considered intervals of more
  than $10$ points. Also, to ensure a fitting interval
  $[s_{\min},s_{\max}]$ within the asymptotic region \cite{tau} of
  averaged distributions, we limited $s_{min}$ to be large enough as
  to overcome crossovers at small-size clusters. In all cases, we compared
  the quality of power-law fits with the corresponding one for
  exponential ones, $P(s)=Ce^{-c s}$, within the obtained interval
  $[s_{\min},s_{\max}]$.}


\begin{table}[h!]
\neww{
\begin{tabular}{|c|c|c|c|c|}
\hline
Density 	& fitted exponent & $s_{min}$ & $s_{max}$	&	\multicolumn{1}{c|}{\begin{tabular}[c|]{@{}c@{}} Norm.Log.Lik.\end{tabular}}	\\ \hline
0.02	&	3.02	&	141		&	5012	&	-0.01484\\	\hline
0.04	&	3.10	&	178		&	5623	&	-0.01046\\	\hline
0.06	&	2.77	&	316		&	17783	&	-0.00506\\	\hline
0.08	&	2.78	&	447		&	14125	&	-0.00278\\	\hline
0.10	&	2.80	&	501		&	22387	&	-0.00144\\	\hline
0.12	&	2.76	&	708		&	25119	&	-0.00153\\	\hline
0.14	&	2.73	&	794		&	25119	&	-0.00128\\	\hline
0.16	&	2.84	&	708		&	28184	&	-0.00157\\	\hline
0.18	&	2.81	&	1000	&	31623	&	-0.00095\\	\hline
0.20	&	2.88	&	1259	&	39811	&	-0.00063\\	\hline
0.22	&	2.79	&	794		&	25119	&	-0.00086\\	\hline
0.24	&	2.73	&	1413	&	50119	&	-0.00067\\	\hline
0.26	&	2.78	&	1995	&	70795	&	-0.00029\\	\hline
0.28	&	2.79	&	2512	&	89125	&	-0.00026\\	\hline
0.30	&	2.79	&	2239	&	79433	&	-0.00026\\	\hline
0.32	&	2.83	&	2239	&	89125	&	-0.00026\\	\hline
0.34	&	2.76	&	3548	&	112202	&	-0.00034\\	\hline
0.36	&	2.84	&	3981	&	141254	&	-0.00015\\	\hline
0.38	&	2.80	&	4467	&	141254	&	-0.00013\\	\hline
0.40	&	2.87	&	3981	&	177828	&	-0.00019\\	\hline
0.42	&	2.81	&	2818	&	89125	&	-0.00010\\	\hline
0.44	&	2.79	&	5012	&	158489	&	-0.00020\\	\hline
0.46	&	2.75	&	5012	&	158489	&	-0.00049\\	\hline
0.48	&	2.59	&	5012	&	223872	&	-0.00184\\	\hline
0.50	&	2.31	&	5012	&	158489	&	-0.00197\\	\hline
0.52	&	2.33	&	5012	&	158489	&	-0.00077\\	\hline
0.54	&	2.35	&	7079	&	281838	&	-0.00016\\	\hline
0.56	&	2.31	&	6310	&	316228	&	-0.00021\\	\hline
0.58	&	2.01	&	3162	&	63096	&	-0.29824\\	\hline
0.60	&	3.22	&	447		&	25119	&	-0.00462\\	\hline
0.62	&	3.72	&	316		&	10000	&	-0.00248\\	\hline
0.64	&	3.70	&	141		&	8913	&	-0.01496\\	\hline
0.66	&	3.74	&	141		&	2818	&	-0.01072\\	\hline
0.68	&	3.53	&	141		&	3162	&	-0.01011\\	\hline
0.70	&	3.53	&	112		&	3162	&	-0.00805\\	\hline
\end{tabular}
}
\caption{{\bf{Fits for averaged cluster-size distributions
      $P(s)$ with different densities.}} Power-laws have been fitted
  (see Appendix) to  averaged $P(s)$ for different densities (first
  column). We show the fitted exponents (second column), the optimal
  interval of clusters sizes --$s_{min}$ (third column) to $s_{max}$
  (fifth column)-- and the corresponding normalized log-likelihood
  for each density.}
\label{table:fit_averaged}
\end{table}

\begin{table}[h]
\begin{tabular}{|c|c|c|c|c|}
\hline
Density 	&
           \multicolumn{1}{c|}{\begin{tabular}[c|]{@{}c@{}} $\overline{\tau}$\\
                                 exponent\end{tabular}} 	&
                                                           \multicolumn{1}{c|}{\begin{tabular}[c|]{@{}c@{}}
                                                                                 Exponent
                                                                                 \\
                                                                                 variance \end{tabular}}
                & \multicolumn{1}{c|}{\begin{tabular}[c|]{@{}c@{}}$\%$
                                        power law better \\than
                                        exponential\end{tabular}}  &
                                                                     \multicolumn{1}{c|}{\begin{tabular}[c|]{@{}c@{}}$\%$
                                                                                           more
                                                                                           than
                                                                                           1.5\\ decades fit\end{tabular}} \\ \hline
0.04	&	2.14	&	1.10	&	71.74	&	71.74	\\ \hline
0.06	&	2.38	&	1.26	&	82.54	&	61.90	\\ \hline
0.08	&	2.46	&	1.21	&	81.48	&	59.26	\\ \hline
0.10	&	2.63	&	1.33	&	78.95	&	40.00	\\ \hline
0.12	&	2.67	&	1.42	&	70.16	&	32.26	\\ \hline
0.14	&	2.75	&	1.32	&	64.29	&	26.62	\\ \hline
0.16	&	2.79	&	1.30	&	62.23	&	31.38	\\ \hline
0.18	&	2.70	&	1.34	&	62.00	&	28.50	\\ \hline
0.20	&	2.57	&	1.24	&	73.50	&	32.50	\\ \hline
0.22	&	2.49	&	1.24	&	81.00	&	43.00	\\ \hline
0.24	&	2.35	&	1.13	&	93.00	&	37.00	\\ \hline
0.26	&	2.43	&	1.36	&	100.00	&	45.50	\\ \hline
0.28	&	2.33	&	1.24	&	100.00	&	55.00	\\ \hline
0.30	&	2.16	&	1.05	&	100.00	&	54.00	\\ \hline
0.32	&	2.05	&	0.89	&	100.00	&	60.00	\\ \hline
0.34	&	1.93	&	0.63	&	100.00	&	66.50	\\ \hline
0.36	&	1.80	&	0.37	&	100.00	&	70.00	\\ \hline
0.38	&	1.71	&	0.23	&	100.00	&	75.50	\\ \hline
0.40	&	1.62	&	0.11	&	100.00	&	84.00	\\ \hline
0.42	&	1.58	&	0.07	&	100.00	&	87.50	\\ \hline
0.44	&	1.60	&	0.07	&	100.00	&	92.00	\\ \hline
0.46	&	1.62	&	0.06	&	100.00	&	94.97	\\ \hline
0.48	&	1.66	&	0.07	&	100.00	&	94.95	\\ \hline
0.50	&	1.70	&	0.06	&	100.00	&	96.97	\\ \hline
0.52	&	1.77	&	0.07	&	100.00	&	96.48	\\ \hline
0.54	&	1.86	&	0.11	&	99.49	&	95.43	\\ \hline
0.56	&	1.96	&	0.18	&	100.00	&	93.91	\\ \hline
0.58	&	2.01	&	0.15	&	100.00	&	92.89	\\ \hline
0.60	&	2.10	&	0.18	&	100.00	&	89.23	\\ \hline
0.62	&	2.15	&	0.23	&	99.48	&	88.02	\\ \hline
0.64	&	2.24	&	0.32	&	98.97	&	63.92	\\ \hline
0.66	&	2.39	&	0.57	&	99.47	&	61.38	\\ \hline
0.68	&	2.50	&	0.71	&	97.30	&	80.54	\\ \hline
\end{tabular}
\caption{{\bf{Fits for individual snapshots with different
      densities.}} We fit power-law curves (see Appendix) for
  snapshots of the model with temporal variability for different
  densities (as listed in the first column). The second column shows
  the mean fitted exponents $\overline{\tau}(\rho)$; the third column shows the variance of
  exponent values (only snapshots whose optimal range considers more
  than $10$ points are considered); the fourth one shows the
  percentage of snapshots for which power-law fits are better than
  exponential ones (i.e. power-laws give a higher normalized
  log-likelihood). The fifth column shows the fraction for which the
  power-law fit extends for at least $1.5$ decades.}
\label{table:fit_snapshots}
\end{table}

\section*{Acknowledgements} 

We acknowledge the Spanish Ministry and Agencia Estatal de
investigaci{\'o}n (AEI) through grant FIS2017-84256-P (European
Regional Development Fund), as well as the Consejer{\'\i}a de
Conocimiento, Investigaci{\'o}n y Universidad, Junta de Andaluc{\'\i}a
and ERDF, Ref.  A-FQM-175-UGR18 and SOMM17/6105/UGR and for
financial support.  We thank Juan A. Bonachela, Paolo Moretti, Victor
Buend{\'\i}a, Guillermo Barrios, Johannnes Zierenberg, and Pablo
Villegas for a critical reading of the manuscript as well as for very
useful comments.

\section*{References}

\bibliographystyle{abbrvnat}
\def\url#1{}

\end{document}